\documentclass[angeo, manuscript]{copernicus}

\usepackage{amssymb} \usepackage{amsmath} \usepackage{bm} \usepackage[percent]{overpic}
\usepackage{longtable} \usepackage{pdflscape}

\ifdefined\pdfsuppresswarningpagegroup
\fi

\AtBeginDocument{\nolinenumbers}

\begin{document}

\title{Optical and Radar Observations of the February 2025 Falcon 9 Upper-Stage Re-entry}

\Author[1,2]{Juha}{Vierinen}
\Author[1]{Dabrowka}{Knach}
\Author[2]{Jorge L.}{Chau}
\Author[2]{Gerd}{Baumgarten}
\Author[2]{Devin}{Huyghebaert}
\Author[2]{Matthias}{Clahsen}
\Author[2]{Nico}{Pfeffer}
\Author[2]{Toralf}{Renkwitz}
\Author[2]{Robin}{Wing}
\Author[3]{Kenneth S.}{Obenberger}
\Author[1]{Björn}{Gustavsson}
\Author[4]{Daniel}{Kastinen}

\affil[1]{Department of Physics and Technology, UiT The Arctic University of Norway, Troms\o, Norway}
\affil[2]{Leibniz Institute of Atmospheric Physics at the University of Rostock, Kühlungsborn, Germany}
\affil[3]{Space Warfare Directorate, Air Force Research Laboratory, Kirtland Air Force Base, New Mexico, USA}
\affil[4]{Swedish Institute of Space Physics, Kiruna, Sweden}

\correspondence{Juha Vierinen}

\runningtitle{Optical and Radar Observations of Falcon 9 Re-entry}
\runningauthor{Vierinen et al.}

\received{}
\pubdiscuss{}
\revised{}
\accepted{}
\published{}

\firstpage{1}

\maketitle

\begin{abstract}
We investigate the February 19, 2025, re-entry of a Falcon 9 upper stage using optical observations from 43 meteor cameras across central Europe together with radar detections of re-entry plasma obtained with the 32.55 MHz SIMONe Germany multistatic radar system. Optical observations of fragment emissions between 85 and 36 km altitude were used to reconstruct 30 fragment trajectories, identify two main fragment families, and fit ballistic trajectories to estimate kinetic energy loss per unit mass. The optical detection-height distribution peaks near 60 km with a standard deviation of 10 km, and both optical and radar signatures occur in the same broad altitude region as the maximum kinetic-energy loss. Radar echoes were detected at altitudes between 55 and 75 km, and the radar-derived positions are consistent with those obtained from optical observations. Two distinct radar echo types associated with the re-entry plasma were identified: (1) specular trail echoes from overdense wake plasma, with radar cross-sections (RCS) of up to 60 dBsm, and (2) short-lived non-specular trail echoes with RCS values of 20--30 dBsm, exhibiting a delay of 1--2 s compared to optical signatures. The characteristic decay time of both echo types is approximately 1 s. In the radar-echo altitude range, the estimated Knudsen numbers for meter-scale fragments are well below unity, consistent with continuum-flow conditions and shock-driven plasma production rather than ordinary meteor-like impact ionization. These serendipitous radar observations demonstrate that the atmospheric re-entry of other spacecraft, including objects smaller than the Falcon 9 upper stage such as Starlink satellites, may likewise be detectable using comparable multistatic meteor radar systems deployed globally.
\end{abstract}
\introduction
\label{sec1}

The rapid commercialization of space has transformed Earth's orbital environment. Rising launch
rates, large satellite constellations, and expanded ride‑share missions have substantially
increased the number, mass, and surface area of objects in low Earth orbit (LEO). Many of these
objects orbit at altitudes where atmospheric drag causes orbital decay within years.
Projected growth in space activity could increase the annual rate of atmospheric re‑entries by a
factor of 2 to 3 over the next decade \citep{report25}, creating an urgent need for improved
understanding of re‑entry processes, deposition of spacecraft materials impacting atmospheric
chemistry and radiative balance, and ground safety risks.

A significant fraction of the mass re-entering the atmosphere originates from spent upper stages of
launch vehicles. Falcon 9 and Falcon Heavy upper stages alone accounted for 0.36 kt out of 1.59 kt
of spacecraft mass influx to the top of the atmosphere in 2024 \citep{schulz26}, making observations
of re-entries of these objects valuable, as they represent a significant portion of the re-entry
flux. This influx is increasingly important to characterize because ablated spacecraft material can
introduce anthropogenic metals into the middle atmosphere, where recent observational and modeling
studies suggest they may influence aerosol composition and stratospheric chemistry
\citep{murphy2023metals,wing2025measurement,maloney2025reentry}. Although the total annual
spacecraft influx of about 1.6~kt corresponds to only about 400~ppt of the total atmospheric
mass between 50 and 70~km, its composition rather than its bulk mass may still be important if
these materials have unusually high chemical reactivity, catalytic activity, efficient radiative
absorption, or unusually long atmospheric lifetimes, many of which remain poorly constrained at
present.


 


In the early hours of 19 February 2025, a Falcon 9 upper stage from the Starlink 11-4 mission
underwent an atmospheric re-entry over central Europe. The Polish Space Agency monitored the object
because of its orbital inclination of approximately 53$^\circ$, and predicted that the re-entry
trajectory could pass over Poland, posing potential ground hazards.
After the event, nine fragments were recovered on the ground and characterized along a strewn field
spanning approximately 70~km \citep{kruzynski2025february}.

This event represents a unique opportunity to investigate spacecraft re-entry processes using a
multi-instrument dataset. Observations are available from optical all-sky fireball cameras
\citep{hankey2020all}, specular radars capable of detecting re-entry plasma
\citep{chau2022atmospheric}, infrasound sensors \citep{hupe2025seismoacoustic}, as well as resonance
lidar observations of Lithium \citep{wing2025measurement}.

Plasma sheaths around re-entry vehicles have long been studied because they can attenuate or interrupt communications with the ground. Classical studies analyzed radio transmission through lifting-vehicle plasma sheaths \citep{evans1963radio} and used Langmuir probes to estimate electron
densities surrounding re-entry vehicles \citep{scharfman1965langmuir}. The same ionized flow is also
relevant for radar phenomenology, because  plasma-clad
bodies can exhibit modified scattering behavior \citep{miller1966scattering,mitchell1969modeling}.
While theoretical work exists on radar scattering from sheath plasma for hypervelocity entry
vehicles \citep{mitchell1969modeling,cong2022numerical,qian2017modelling,esposito2024non},
observations of re-entering orbital spacecraft remain scarce in the literature. Notable examples are
the radio echoes reported during the Mercury MA-6 re-entry \citep{lin1962radio}, airborne radar
measurements obtained during the Ariane 5 EPC re-entry \citep{rubin2005airborne}, and laboratory
plasma measurements performed by \citep{petervari2026radar}.

On the other hand, a significant body of work exists on observations and modeling of radar echoes
associated with meteors. It is known that the plasma created during hypervelocity atmospheric entry
of meteors can substantially enhance the radar detectability of the meteoroids, producing meteor head
echoes \citep{close2007meteor} that are associated with the plasma surrounding the ablating meteoroid,
and specular trail echoes that have very large radar cross-sections (RCS) due to the coherent
addition of scattered EM waves along an extended Fresnel zone
\citep{mckinley1961meteor,poulter1977radiowave}. These radar echoes from meteoric plasma make it
possible to detect micrometeoroids entering the Earth's atmosphere using radar, which would not be sensitive enough to observe echoes from small meteoroid bodies without the significantly larger plasma structure surrounding it.

The objectives of this study are to: (1) reconstruct the three-dimensional trajectories of re-entry
fragments using all-sky camera triangulation, (2) fit ballistic trajectories to estimate re-entry
dynamics to allow analysis of radar observations, and (3) characterize the radar signatures
associated with the re-entry debris.



\section{Instrumentation and Data} \label{sec:data}

\begin{figure}[h!]
    \centering
    \includegraphics[width=\linewidth]{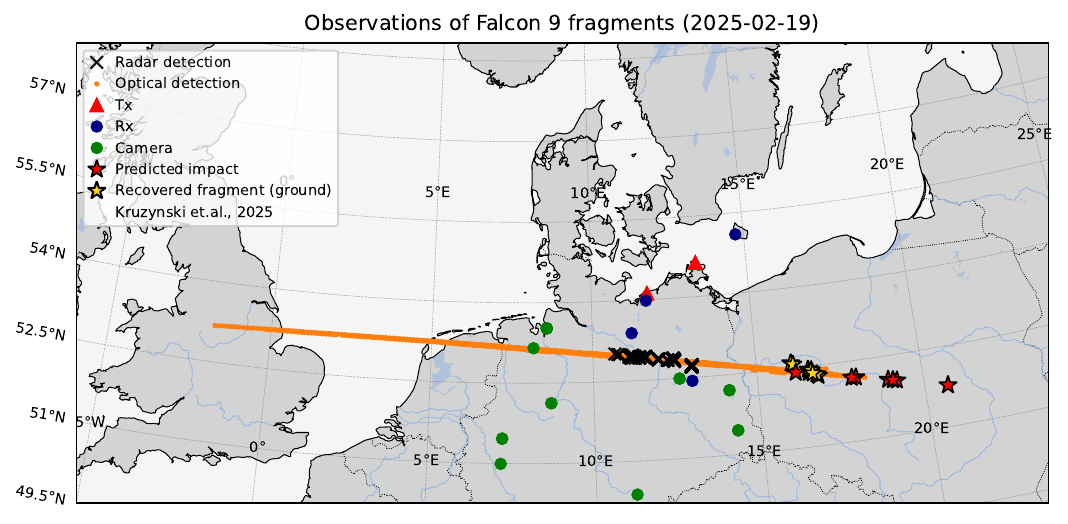}
    \caption{Map of the study area showing SIMONe radar transmitters (triangles), SIMONe radar
             receivers (circles), optical fragment detections (orange), selected radar detections (black
             crosses), recovered ground fragments reported by \cite{kruzynski2025february} in Poland
             (yellow stars), and  projected ground fragment locations extrapolated from
             optical trajectories (red stars). The AllSky7 cameras used in
             the triangulation are shown with green circles.}
    \label{fig:map_falcon9}
\end{figure}

The two datasets used in this study are optical observations from the AllSky7 meteor camera network
\citep{hankey2020all} and radar observations from the SIMONe (Spread Spectrum Interferometric Multistatic meteor radar Observing Network) Germany multistatic radar system
\citep{chau2022atmospheric}. Together, they provide complementary information on the re-entry, with
the cameras tracking the luminous fragment trajectories and the radar probing the associated plasma
signatures. An overview of the observing geometry, detections, recovered ground fragments in Poland
reported by \cite{kruzynski2025february}, and qualitative projected impact locations is given in
\autoref{fig:map_falcon9}.

\subsection{AllSky7 Meteor Camera Network}

Fireball camera networks have already shown their value for reconstructing spacecraft re-entries,
constraining upper-stage orbits from video observations, and investigating upper-stage
fragmentation during re-entry \citep{pena2021using,bartkova2023investigation}. The optical
observations analyzed in this study were acquired through the European AllSky7 Fireball Network.
Each station consists of a camera system equipped with seven NetSurveillance NVT cameras with the
Sony Starvis IMX291 sensor and a 4\,mm f/1.0 lens, recording at 25\,fps. Each camera has a field of
view of about $40^\circ \times 85^\circ$. To cover the full sky, five of the cameras are
horizontally oriented at an inclination of $\sim 25\degree$, while the remaining two cameras cover
the northern and southern directions at an inclination of $\sim 70\degree$ \citep{hankey2020all}.

 \subsection{SIMONe Germany}

 The radar data analyzed in this work were collected with the SIMONe Germany system, a multistatic specular meteor radar system developed to observe atmospheric winds in the
 mesosphere and lower thermosphere \citep{chauetal2019, vierinen2019observing}
 using underdense specular meteor trail echoes with diverse viewing angles. The network comprises two transmitting stations, located in Juliusruh
 and Kühlungsborn. Each site operates six transmitting antennas that emit independent 1000-bit
 pseudorandom binary phase-coded continuous-wave signals \citep{vierinen2016coded,chauetal2019} at 32.55 MHz, with
 a combined transmit power of 3 kW (500 W per antenna). This setup
 permits interferometric direction finding using single receiver antennas, provides an unaliased total
 propagation range of 3000 km, and supports an unaliased Doppler shift range of $\pm 50$ kHz. The positions of the transmitters in Juliusruh and Kühlungsborn and the receivers used in this work, i.e., in Bornholm, Bornim,
 Moitin, and Hagenow, are illustrated in \autoref{fig:map_falcon9}.

\section{Methods}

The AllSky7 meteor camera network measurements were triangulated using dual camera observations to
derive the positions of the fragments as a function of time. A ballistic fit of the trajectories was
performed to allow the calculation of the energy loss rate and vector velocity along the fragment paths. The
SIMONe radar observations were processed to identify the location, Doppler shift, and RCS of the
echoes associated with the re-entry debris. The custom data-processing and figure-generation
software used in this study is publicly available \citep{falcon9_github}.

\subsection{Dual camera triangulation}

\begin{figure}[h!]
    \centering
    \includegraphics[width=0.5\linewidth]{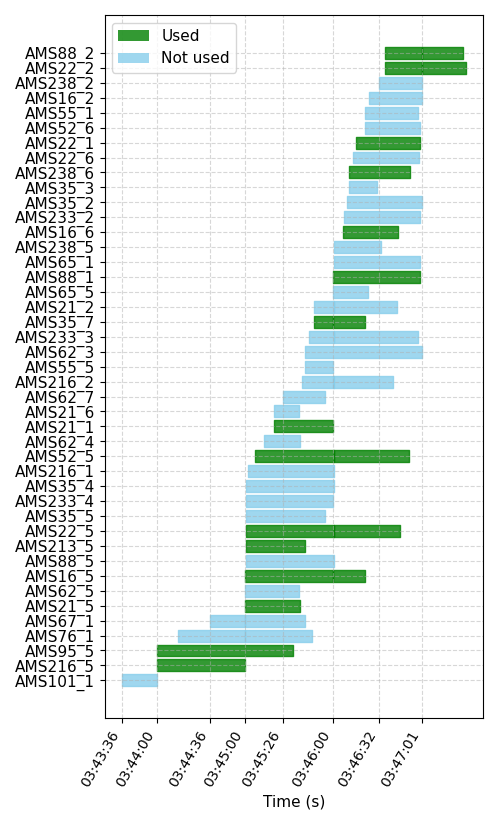}
    \caption{Visibility intervals of the fragments as seen from the available AllSky7 stations, where the underscored number indicate the per station camera id. This
             overview helped identify compatible observation pairs for triangulation and ensured
             coverage of all time segments.}
    \label{fig:visibility}
\end{figure}

 The optical data processing focused on two main tasks: star calibration and stereoscopic triangulation of
 fragment positions. The February 19th event was primarily observed from stations in Germany, but
 one observation from the United Kingdom (AMS101) and two observations from Belgium (AMS67) are also
 available. \autoref{fig:visibility} shows the time and duration of fragment visibility in each
 available observation. Pairs of overlapping videos chosen for calibration and triangulation were
 selected based on video quality, duration, fragment visibility, overlap with other observations,
 and star visibility. Out of 43 videos, 15 were calibrated and included for further stereoscopic
 triangulation. Although AMS101 detected the re-entry at 03:43:00 UTC, no stars were visible
 in that video to allow calibration, and no additional cameras were available for triangulation at
 that time.


An average of about 100 stars was identified in most calibrated camera observations with good
fragment visibility, enabling accurate plate solving and lens modeling. Star calibration was
performed on the same video that contains the re-entry to minimize calibration errors due to camera
movement, and star positions were cross-checked during triangulation to detect timing offsets or
camera motion. The geometric camera calibration and triangulation were carried out with AIDA\_tools
\citep{aida_tools_github} and followed the procedures described by \cite{gustavsson2008first} and
\cite{vierinen2022multi}. The camera lens-model calibration residuals were around $\pm
1/4$ pixel.

\begin{figure*}[h!]
    \centering
    \setlength{\tabcolsep}{2pt}
    \renewcommand{\arraystretch}{0}
    \begin{tabular}{@{}cc@{}}
        \begin{overpic}[width=0.42\textwidth]{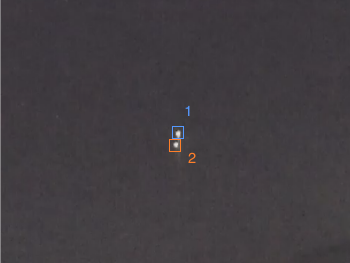}
            \put(3,3){\color{white}\large a)}
        \end{overpic}
        \begin{overpic}[width=0.42\textwidth]{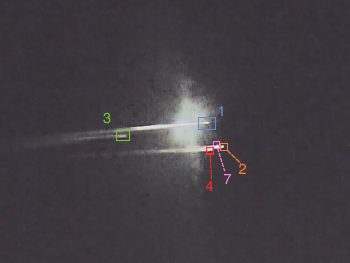}
            \put(3,3){\color{white}\large b)}
        \end{overpic}\\
        \begin{overpic}[width=0.42\textwidth]{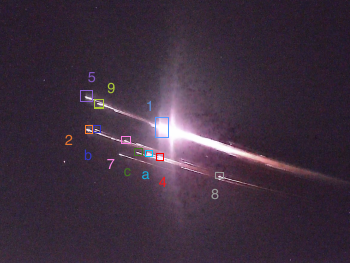}
            \put(3,3){\color{white}\large c)}
        \end{overpic}
        \begin{overpic}[width=0.42\textwidth]{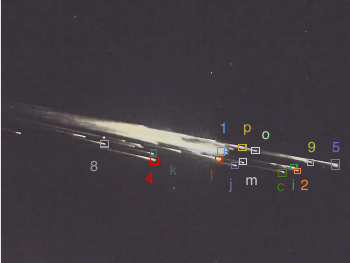}
            \put(3,3){\color{white}\large d)}
        \end{overpic}\\
        \begin{overpic}[width=0.42\textwidth]{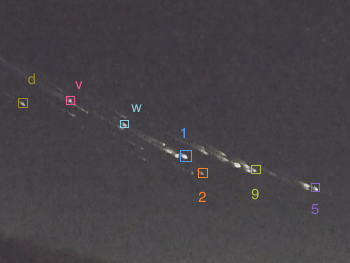}
            \put(3,3){\color{white}\large e)}
        \end{overpic}
        \begin{overpic}[width=0.42\textwidth]{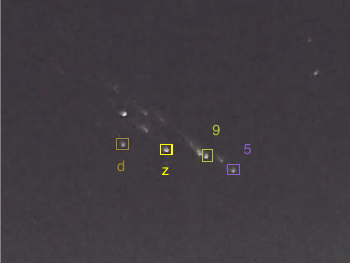}
            \put(3,3){\color{white}\large f)}
        \end{overpic}
    \end{tabular}
    \caption{Six representative AllSky7 frames of the re-entry debris, ordered by time. Panels a--f
             show AMS95 at 03:44:30, AMS21 at 03:45:30, AMS35 at 03:46:02, AMS16 at 03:46:20, AMS88
             at 03:47:00, and AMS22 at 03:47:15. Fragment identifiers for triangulated fragments are
             shown with colored numbers and letters. Panels c) and d) are during the time when radar
             echoes of the re-entry plasma were observed (3:46:00-3:46:30). All times are UTC.}
    \label{fig:grid_2x3}
\end{figure*}

\autoref{fig:grid_2x3} shows six representative frames from six different cameras, ordered
chronologically through the re-entry event. Together, the panels illustrate the evolution of the
fragmentation from the early phase of the breakup to the final visible fragments near the end of the
event. Assigned fragment identifiers are labeled on the plot.

Fragments were located and triangulated using pairs of calibrated videos. Videos recorded at 25\,fps
were sampled at 1\,s intervals (every 25th frame) for the triangulation runs. Detected fragments
were assigned identifiers using the ranges 1–9 and a–z. For each detection, we stored an entry
in Hierarchical Data Format (HDF5) containing the fragment ID, position in Earth-centered
Earth-fixed (ECEF) coordinates, geographic coordinates, observation time, estimated altitude, and
position uncertainty. As a metric of position uncertainty, we used the shortest distance between
the two line-of-sight vectors from each camera. The estimated position uncertainties were 150 m and
480 m at the 67\% and 95\% percentiles, respectively.

One practical lesson learned for future optical observations is that accurate absolute timing is
essential for fragment identification and triangulation during complex break-up events. The timing
for the videos is provided by the network timing protocol, but the timing offset between the video
frame timestamp and the time of exposure is not well known. Using a light emitting diode (LED)
blinker synchronized to global positioning system (GPS) time, we determined that AllSky7 timing is
at most 0.3 seconds offset from GPS time \citep{kastinen2025preparing}. One desirable modification
to existing AllSky7 meteor cameras would be to incorporate an external optical timing reference, for
example, with GPS-synchronized LED blinkers or similar timing fiducials visible in the camera
frames, or alternatively, modifying the camera firmware to accept a time-stamping signal directly from a local GPS receiver or Network Time Protocol (NTP) server.

\subsection{SIMONe Radar Measurements}

In this work, we present the radar echoes detected with the following SIMONe Germany transmit-receive links: Juliusruh-Bornim, Juliusruh-Bornholm, Juliusruh-Hagenow, Juliusruh-Moitin, Kühlungsborn-Bornim,
Kühlungsborn-Bornholm, Kühlungsborn-Hagenow, and Kühlungsborn-Moitin.
Interferometric direction finding, together with transmit-receive delay measurements, provided
estimates of the scattering location, received signal power, RCS, and Doppler shift.


Due to the rapid motion of the re-entry target, the echoes were processed using a range-Doppler
matched filter bank at a 100 kHz Doppler bandwidth, with a 10 ms coherent integration window. This
resulted in a $B_{\mathrm{rx}}=100$ Hz receiver noise bandwidth. This allowed for the detection of large
Doppler shifts, typically associated with meteor head echoes. For details on the processing steps,
refer to \cite{chau2021multistatic}, which discusses observations of fast-moving meteor head echoes
using the same technique but with the SIMONe Peru system. Similar processing was also used to estimate the range and Doppler shift for a head echo associated with a fireball with SIMONe New Mexico \citep{obenberger2026first}.


Accurate Doppler shift, signal-to-noise ratio, and angle-of-departure \citep{chau2019phasecal} estimates were obtained using signals that are decoded with a compressive sensing approach based on
\cite{urco2019sparse}, which exploits the range sparsity of the observed echoes. The decoding
consists of an iterative detection and removal of strong, medium, and weak echoes, followed by a
final least squares inversion over all detected signals. We extended this approach to jointly handle
multiple transmit sites within each iteration, where the detection is performed separately for each
site before jointly subtracting all contributions from the measurements. This reduces the
cross-interference between the two SIMONe Germany transmit sites in Kühlungsborn and Juliusruh,
which transmit at the same frequency using pseudorandom codes. 



The bistatic RCS was estimated from the measured signal-plus-noise to noise ratio,
\begin{align}
\eta = \frac{S+N}{N},
\end{align}
where $\eta$ is expressed in linear units. 
Assuming receiver noise power
\begin{align}
N = k_B T_\mathrm{noise} B_\mathrm{rx},
\end{align}
the received signal power is
\begin{align}
S = (\eta - 1) N.
\end{align}
The bistatic RCS is then estimated using
\begin{align}
\sigma_b = \frac{S (4\pi)^3 R_\mathrm{tx}^2 R_\mathrm{rx}^2}
{P_\mathrm{tx} G_\mathrm{tx} G_\mathrm{rx} \lambda^2},
\end{align}
where $R_\mathrm{tx}$ and $R_\mathrm{rx}$ are the transmitter-to-target and target-to-receiver
ranges, $P_\mathrm{tx}$ is the transmit power, $G_\mathrm{tx}$ and $G_\mathrm{rx}$ are the transmit
and receive antenna gains, and $\lambda$ is the radar wavelength.

In the present analysis, the conversion was carried out with the fixed parameters used in the
processing code: $f=32.55$ MHz, $P_\mathrm{tx}=500$ W, $G_\mathrm{tx}=G_\mathrm{rx}=1$,
$B_\mathrm{rx}=100$ Hz, and $T_\mathrm{noise}=6000$ K. Although six transmitters were used, their contributions have been added incoherently. The estimated RCS values are based on a
simplified radar equation, and they neglect additional effects such as polarization mismatch,
detailed antenna pattern variations, and other propagation or processing losses. The values are not
expected to be wrong by significantly more than a factor of 16.

\subsection{Dynamic modeling of re-entry}
\label{sec:dynamic_model}

For the ballistic trajectory fitting, we used a simple point-mass model in Earth-centered inertial
(ECI) coordinates, i.e., fixed relative to the stars. Optical positions triangulated in ECEF coordinates were transformed to ECI coordinates. The fitted state obeys
\begin{align}
    \frac{d\bm{v}}{dt} &= \bm{a}_{g}(\bm{r}) - \frac{1}{2}\rho_a(\bm{r},t)\, B(t)\lVert \bm{v}_{\mathrm{rel}} \rVert \bm{v}_{\mathrm{rel}},
    \label{eq:dynamic_model}
\end{align}
where $\bm{r}$ is the fragment position vector in ECI coordinates, $\bm{a}_{g}$ is the gravitational
acceleration including the Earth's $J_2$ term, and $\rho_a$ is the atmospheric density evaluated
using the NRLMSIS~2.1 model \citep{emmert2022nrlmsis}. The drag is based on the velocity relative to
the neutral atmosphere, $\bm{v}_{\mathrm{rel}} = \bm{v} - \bm{v}_{\mathrm{atm}}$, where
$\bm{v}_{\mathrm{atm}}$ is the atmospheric velocity in the inertial frame. The factor $B(t)$ denotes
the effective drag-to-mass parameter,
\begin{align}
    B(t)=\frac{C_D A \rho_a'}{m},
\end{align}
where $C_D$ is the drag coefficient, $A$ is the effective cross-sectional area, $\rho_a'$ is a
relative correction to the atmospheric density accounting for the mismatch between the actual
neutral density and the model density, and $m$ is the fragment mass. All four of these terms in the
ballistic coefficient are expected to vary during re-entry. To allow slow changes in drag behaviour
during fragmentation, the ballistic coefficient was parameterized in log-space as $B(t) =
10^{\ell(t)}$, where $\ell(t)$ was linearly interpolated between four fitted node values over the
observation interval.

The free parameters were therefore the initial ECI position $\bm{r}_0$, initial ECI velocity
$\bm{v}_0$, and the four fitted node values defining $\ell(t)$. These parameters were estimated by
minimizing the sum of squared position residuals,
\begin{align}
    S = \sum_{i=1}^{N} \left\lVert \bm{r}_{\mathrm{mod}}(t_i) - \bm{r}_{\mathrm{obs}}(t_i) \right\rVert^2,
\end{align}
using the triangulated optical positions $\bm{r}_{\mathrm{obs}}(t_i)$. The initial guess was formed
from the first triangulated position and a finite-difference velocity estimate from the first and
last optical samples, and the model $\bm{r}_{\mathrm{mod}}(t)$ was integrated numerically with a
first-order Euler-Cromer (semi-implicit Euler) scheme using a fixed 0.5\,s time step during the
optimization.

The wind-corrected fits were carried out in two stages. First, a zero-horizontal-wind solution was
obtained assuming atmospheric co-rotation, $\bm{v}_{\mathrm{atm}}=\bm{\Omega}_{\oplus}\times\bm{r}$.
ERA5 reanalysis winds \citep{hersbach2020era5,hersbach2023era5_pressure_levels} were then sampled along this nominal trajectory,
converted to inertial atmospheric velocity $\bm{v}_{\mathrm{atm}}=\bm{\Omega}_{\oplus}\times\bm{r} +
\bm{v}_{\mathrm{ERA5}}$, and interpolated in time for the final optimization. Above the highest
altitude represented by the ERA5 profile, the horizontal wind was set to zero so that only
atmospheric co-rotation contributed to $\bm{v}_{\mathrm{atm}}$.

This simplified model can reproduce smoothly varying ballistic coefficients and velocities along the
re-entry path, but it does not capture abrupt changes in ballistic coefficient associated with
discrete fragmentation events. The results must therefore be treated with caution.

Because the drag model directly determines the ballistic coefficient, it also determines the
specific kinetic-energy loss rate per unit mass. We therefore evaluated
\begin{align}
    \dot{e}_{\mathrm{loss}}
    = -\bm{v}_{\mathrm{rel}} \cdot \bm{a}_{\mathrm{drag}}
    = \frac{1}{2}\rho_a(\bm{r},t) B(t)\lVert \bm{v}_{\mathrm{rel}} \rVert^3,
\end{align}
which provides a convenient diagnostic of where drag-related kinetic-energy dissipation is strongest
without requiring an independent estimate of fragment mass or cross-sectional area. This quantity relies only on velocity and $B(t)$, and therefore allows relative comparisons of energy dissipation along the trajectory even when decoupled estimates of area, mass, and drag coefficient, and neutral density are unobtainable without additional assumptions. 

To better understand the plasma production within the high temperature shock created by the hypervelocity entry giving rise to radar measurements, we also estimate the temperature $T_{post}$ behind a steady normal shock. We use standard formula,
\begin{align}
    T_{post} = T_{pre} \frac{(2 \gamma M_{pre}^2 - (\gamma - 1))((\gamma - 1)M_{pre}^2 + 2)}{(\gamma + 1)^2M_{pre}^2} \label{eq:shock_temp}
\end{align}
that apply to adiabatic flows of a completely perfect fluid
\citep{Ames1953NACA1135,anderson1989hypersonic}. Here $T_{pre}$ is the atmospheric temperature, $M_{pre}$ is the Mach number of the flow before the shock, and $\gamma$ is the specific heat ratio of the gas. The above equation can be derived from the
Rankine-Hugoniot relations, and the estimated post-shock temperature should be treated as a diagnostic of the possibility for the shock to produce enough free electrons to allow radio wave scattering.

\section{Results}  \label{sec:results}

\subsection{Optical fragment trajectories}

\begin{figure}[h!]
    \centering
    \includegraphics[width=\linewidth]{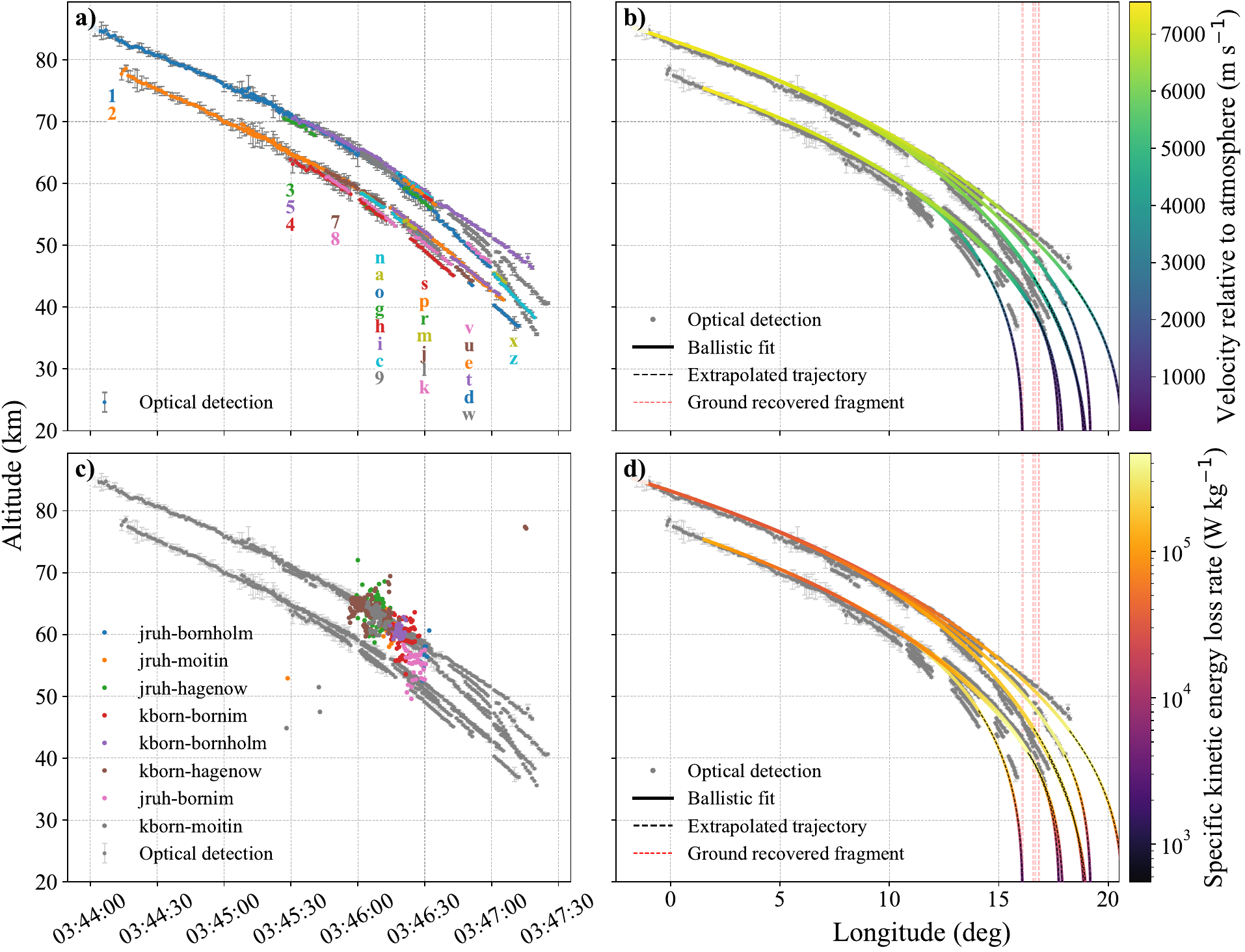}
    \caption{Overview of the optical, dynamical, and radar results in time-altitude and
             longitude--altitude coordinates. Panel a) shows the triangulated optical fragment
             detections with the fragment identifier as a function of time and height. Panel b)
             shows a representative subset of ballistic fits colored by velocity relative to the
             atmosphere (longitude vs height). Panel c) shows the positions of optical fragment and
             radar detections of the re-entry plasma (time vs height). Panel d) shows specific
             kinetic energy loss per unit mass (longitude vs height). Red vertical lines indicate
             longitudes of ground recovered fragments.}
    \label{fig:optical_radar_2x2}
\end{figure}

\autoref{fig:optical_radar_2x2}a shows the altitude of the fragments detected with the meteor
cameras as a function of time. At the time when the first triangulations were possible, one fragment
$F_1$ was visible at an altitude of 85.3 km at 3:44:00 UTC. A second fragment $F_2$ appeared at an
altitude of 78.5 km at 3:44:15. These two fragments continued to fragment further, with the most
intense fragmentation occurring at an altitude of 60 km. Fragments associated with $F_2$ crossed 60
km at approximately 3:45:51, whereas the higher altitude fragment $F_1$ crossed 60 km at
approximately 3:46:22. The last triangulated fragment was observed at an altitude of 40 km at
3:47:25. The lowest triangulation was made for a fragment associated with $F_2$ at an altitude of 36
km at 3:47:12.

\subsection{Ballistic fits and specific energy loss}

\begin{figure}[h!]
    \centering
    \begin{overpic}[width=\linewidth]{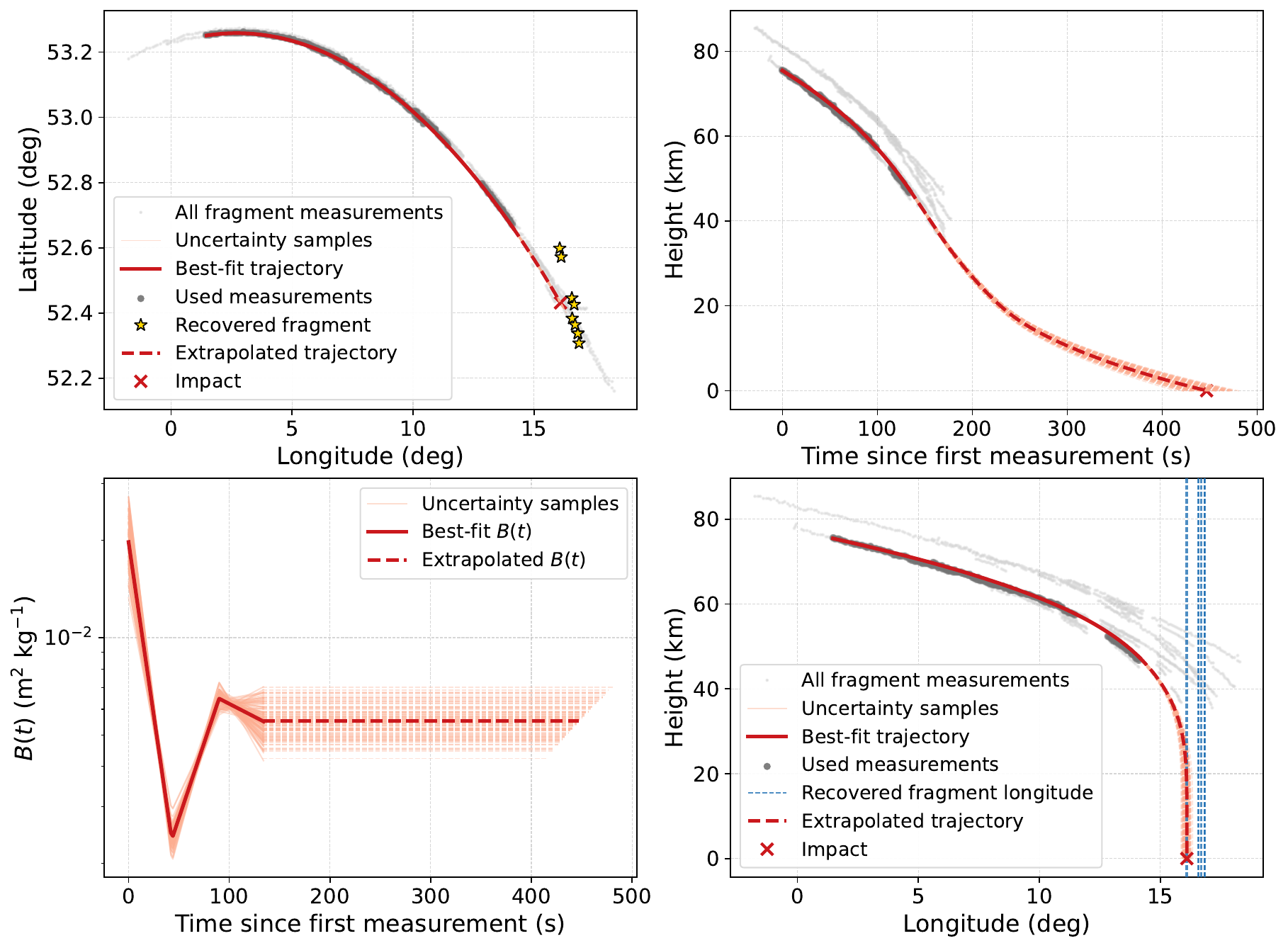}
        \put(3,72){\textbf{a)}}
        \put(53,72){\textbf{b)}}
        \put(3,36){\textbf{c)}}
        \put(53,36){\textbf{d)}}
    \end{overpic}
    \caption{Example ballistic fit for the fragment chain $F_2, F_7, F_a, F_k$. Gray markers show
             all triangulated optical positions, darker gray markers show the measurements used in
             this fit, the solid red curve is the best-fit trajectory, the light red curves show 100
             uncertainty samples, and the dashed red curve shows the extrapolated trajectory toward
             the nominal impact point. Panel a) shows the latitude-longitude ground projection,
             panel b) altitude versus time, panel c) the fitted ballistic coefficient, and panel d)
             longitude versus altitude, together with the recovered fragment locations.}
    \label{fig:ballistic_fit_example}
\end{figure}

Panels b) and d) of \autoref{fig:optical_radar_2x2} show a representative subset of ballistic fits
overlaid on all triangulated optical fragment positions. Not all possible chains of fragmentation
were fit, but the subset shown captures the main behavior of both fragment families and provides a
compact illustration of the inferred dynamical evolution. Panel b) shows that the fitted speeds
relative to the atmosphere start at 7 km\,s$^{-1}$ and are around 3 km\,s$^{-1}$ when optical fragments
disappear. Panel d) shows that the fitted specific kinetic-energy loss rate per unit mass is highest
over a broad altitude range from roughly 30 to 60 km.

The estimates shown in \autoref{fig:optical_radar_2x2}b,d are most reliable within the altitude
interval constrained by the optical measurements and are expected to be most robust near the middle
of that interval, around 60 km altitude. Above and below the observed range, the trajectories,
relative speeds, and energy-loss rates rely increasingly on extrapolation and should therefore be
interpreted with caution.

The extrapolation of the fit for the fragment chain ($F_2, F_7, F_a, F_k$) shown in
\autoref{fig:optical_radar_2x2}b,d reaches the ground near the longitude of one of the recovered
debris fragments. \autoref{fig:ballistic_fit_example} shows the fit for this fragment chain in more
detail, including the fitted ballistic coefficient and the spread of trajectories obtained from the
linearized parameter uncertainty. However, the extrapolated impact locations should not be
interpreted as precise predictions. The ballistic coefficient changes substantially over the course
of fragmentation, and this evolution is difficult to model and fit robustly with the present
parameterization. Therefore, the estimated parameters  carry a significantly larger uncertainty than the linearized uncertainties would indicate. 

A summary of the fitted ballistic coefficients, peak dynamic quantities, and ECEF speeds at selected
altitude levels is given in \autoref{tab:ballistic_fit_summary}. The quoted 2$\sigma$ uncertainties
are derived from the fitted parameter covariance and propagated to the reported quantities. These
values are most informative within the altitude interval constrained by the optical measurements.

\begin{table*}[t]
\centering
\footnotesize
\setlength{\tabcolsep}{4.0pt}
\renewcommand{\arraystretch}{1.12}
\caption{Summary of the fitted fragment chains. $B_0$ and $B_1$ denote the fitted start and end ballistic coefficients in units of $10^{-3}$~m$^2$~kg$^{-1}$. The diameter column gives a rough effective spherical-equivalent size range inferred from $B_0$ and $B_1$ assuming $C_D=2$, an initial bulk density of 30~kg~m$^{-3}$, and a terminal bulk density of 1000~kg~m$^{-3}$. Peak dynamic pressure and peak specific energy-loss rate are reported; uncertainties are estimated as approximate 2$\sigma$ spreads from 50 covariance-drawn trajectory samples. The speed entries give the absolute ECEF speeds at the downward crossings of 70 and 40~km altitude; their uncertainties are estimated as 2$\sigma$ spreads from 50 covariance-drawn trajectory samples.}
\label{tab:ballistic_fit_summary}
\resizebox{\textwidth}{!}{%
\begin{tabular}{lccccccc}
\hline
Path & $B_0$ & $B_1$ & $d_0 \rightarrow d_1$ & $q_{\max}$ & $\dot{e}_{\mathrm{loss},\max}$ & $|\mathbf{v}|_{70}$ & $|\mathbf{v}|_{40}$ \\
Units & $10^{-3}$ m$^2$ kg$^{-1}$ & $10^{-3}$ m$^2$ kg$^{-1}$ & m & kPa & kW~kg$^{-1}$ & km~s$^{-1}$ & km~s$^{-1}$ \\
\hline
$F_{1}$ & $31.6 \pm 9.9$ & $2.6 \pm 0.4$ & $3.16 \rightarrow 1.17$ & $33.3 \pm 5.3$ & $294.9 \pm 3.9$ & $6.90 \pm 0.01$ & $3.93 \pm 0.19$ \\
$F_{2}$ & $24.1 \pm 9.1$ & $3.7 \pm 0.7$ & $4.14 \rightarrow 0.81$ & $26.3 \pm 2.0$ & $373.9 \pm 26.4$ & $7.14 \pm 0.02$ & $3.74 \pm 0.16$ \\
$F_{1}, F_{5}$ & $31.6 \pm 9.9$ & $2.8 \pm 0.2$ & $3.16 \rightarrow 1.08$ & $32.4 \pm 2.1$ & $368.2 \pm 13.6$ & $7.10 \pm 0.01$ & $4.04 \pm 0.10$ \\
$F_{1}, F_{9}$ & $31.6 \pm 9.9$ & $5.2 \pm 0.6$ & $3.16 \rightarrow 0.58$ & $20.2 \pm 1.2$ & $373.0 \pm 13.7$ & $7.00 \pm 0.01$ & $3.18 \pm 0.13$ \\
$F_{2}, F_{7}, F_{a}, F_{k}$ & $24.1 \pm 9.1$ & $5.6 \pm 1.5$ & $4.14 \rightarrow 0.54$ & $12.5 \pm 2.5$ & $276.5 \pm 20.2$ & $7.14 \pm 0.02$ & $2.59 \pm 0.30$ \\
$F_{2}, F_{7}, F_{i}, F_{t}$ & $24.1 \pm 9.1$ & $4.9 \pm 1.0$ & $4.14 \rightarrow 0.61$ & $26.4 \pm 1.5$ & $462.1 \pm 33.8$ & $7.12 \pm 0.02$ & $3.59 \pm 0.21$ \\
$F_{1}, F_{p}, F_{w}, F_{z}$ & $31.6 \pm 9.9$ & $1.8 \pm 0.3$ & $3.16 \rightarrow 1.65$ & $44.3 \pm 6.4$ & $270.7 \pm 5.4$ & $6.91 \pm 0.01$ & $4.10 \pm 0.11$ \\
\hline
\end{tabular}%
}
\end{table*}

\subsection{RCS and Doppler signatures}

\begin{figure}[h!]
    \centering
    \includegraphics[width=\linewidth]{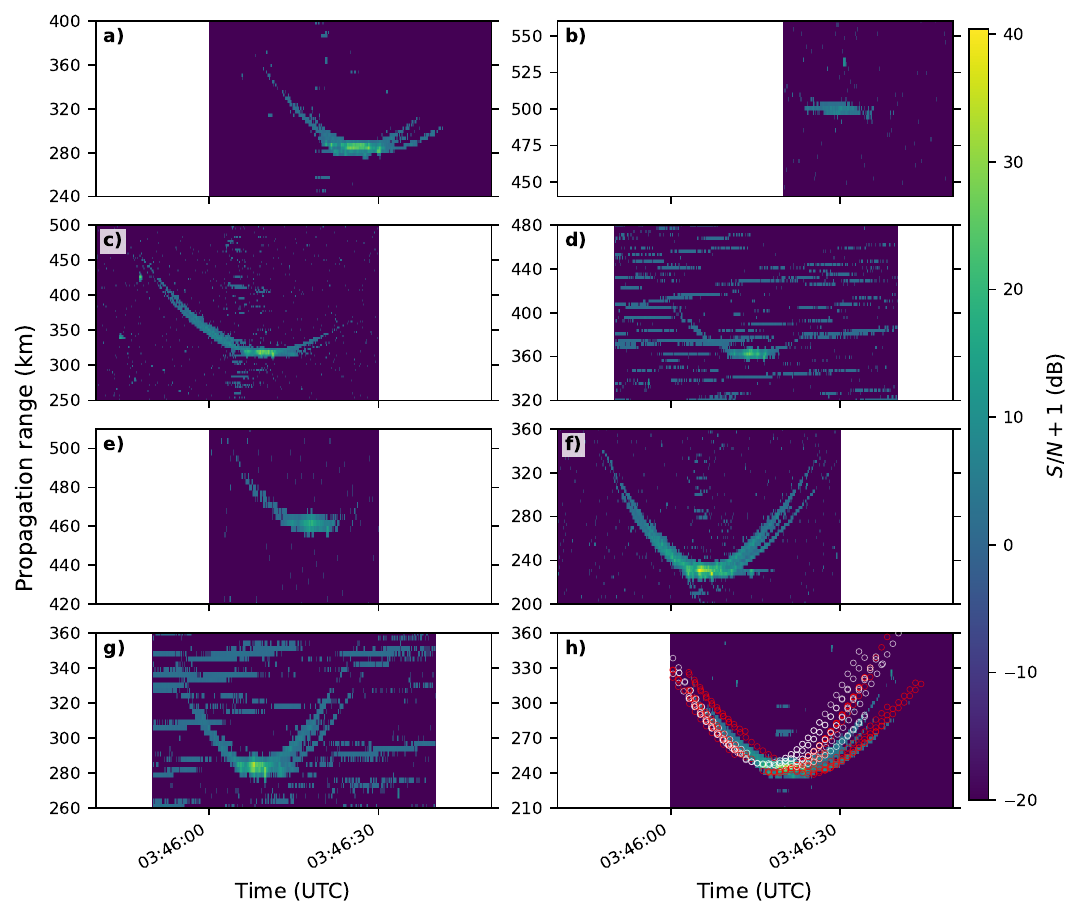}
    \caption{$(S+N)/N$ in dB with 100 Hz noise bandwidth as a function of time and propagation range
             for the eight SIMONe transmitter-receiver links that detected the event. Each panel
             shows one bistatic link with a panel-specific propagation-range interval chosen to show
             the echo region. Optical fragment positions are overlaid only in panel h): open circles
             show the propagation ranges computed from triangulated optical positions after shifting
             the optical times later by 1.5\,s. White circles indicate the $F_1$ fragment family, and
             red circles indicate the $F_2$ fragment family. Both families coincide with the
             non-specular radar echo ranges in this panel.}
    \label{fig:snr_all_links}
\end{figure}

\autoref{fig:snr_all_links} shows the corresponding $(S+N)/N$ measurements for all eight radar links
that detected the event. The radar signature associated with the re-entry plasma appears as a
parabolic arc, first approaching in range, then reaching maximum power near the closest point
corresponding to specular geometry, and finally receding. The observed range interval and temporal
evolution of this radar echo differ between links due to variations in transmitter–receiver
geometry relative to the re-entry trajectory. In panel h), the open circles show the bistatic
propagation ranges expected from the triangulated optical fragment positions after shifting the
optical times later by 1.5\,s. White circles denote fragments associated with the $F_1$ family, and red
circles denote fragments associated with the $F_2$ family. Both fragment families align with the
range-time locations of the non-specular radar echoes in this link.

\begin{figure}[h!]
    \centering
    \includegraphics[width=0.6\linewidth]{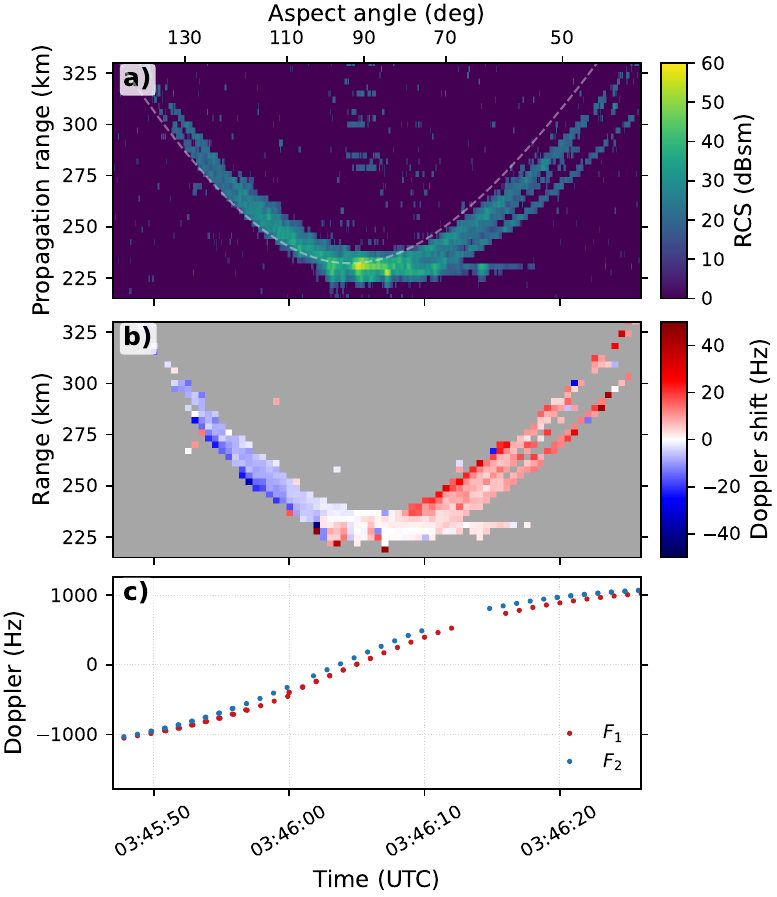}    
    \caption{Radar signatures for the Kühlungsborn-Hagenow link. Panel a) shows the estimated RCS
             as a function of time and propagation range. Panel b) shows the Doppler shift. Panel c)
             shows the Doppler shift predicted from the fitted trajectories of fragments $F_1$ and
             $F_2$.}
    \label{fig:rcs}
\end{figure}

\autoref{fig:rcs}a) shows the estimated RCS measured using the Kühlungsborn-Hagenow link as a
function of time and bi-static propagation range (the sum of the distance from the transmitter to
the target and from the target to the receiver). The predicted full propagation range based on the
fitted trajectory of fragment $F_1$ is shown with a white dashed line, which appears to be shifted
by 1-2 seconds earlier than the similar parabolic arc painted by the radar echoes.

The top axis of \autoref{fig:rcs}a) shows the aspect angle based on the fitted trajectory of
fragment $F_1$:
\begin{equation}
\theta = \arccos\left(\frac{\bm{k}_B \cdot \bm{v}}{\lVert \bm{k}_B \rVert \, \lVert \bm{v} \rVert}\right)
\end{equation}
where $\bm{k}_B$ is the bi-static Bragg wave vector and $\bm{v}$ is the velocity vector of the
scattering re-entry fragment. The peak RCS is observed near the specular aspect angle
(90$^{\circ}$). There are also several other localized enhancements in RCS, which might be
associated with other fragments passing through the specular aspect angle at different times.


\autoref{fig:rcs}b) shows the Doppler shift based on the range-Doppler matched filter bank for the
Kühlungsborn-Hagenow link. The colors in the plot are restricted to $\pm 100$ Hz because no larger
values were estimated. Note that the minimum and maximum values in the pass band of the analysis are
$\pm 50$ kHz. \autoref{fig:rcs}c) shows the expected Doppler shift for this radar link based on the
motion of fragments $F_1$ and $F_2$ with the assumption that the scattering plasma moves at the same
velocity as the parent fragment, as would be the case with a meteor head echo
\citep{close2007meteor}.

\begin{figure}[h!]
    \centering
    \includegraphics[width=0.8\linewidth]{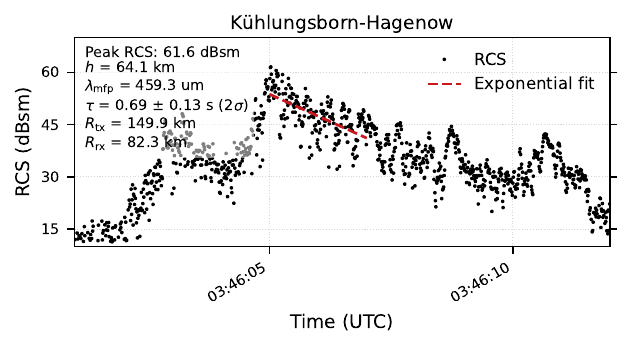}
    \caption{Estimated RCS as a function of time for the Kühlungsborn-Hagenow range gate with the
             highest $(S+N)/N$. The conversion from measured power to RCS uses the
             transmitter-to-target and target-to-receiver ranges estimated from the fitted
             trajectory of fragment $F_1$. The dashed line shows an exponential fit over the
             interval 03:46:05--03:46:07 UTC.}
    \label{fig:rcs_peak_gate}
\end{figure}

To illustrate the temporal evolution of the strongest echo in more detail,
\autoref{fig:rcs_peak_gate} shows the estimated RCS for the Kühlungsborn-Hagenow link at the range
gate with the highest echo power. The time series was evaluated at a time resolution of 10 ms and
converted to RCS using a 100 Hz noise bandwidth together with the transmitter-to-target and
target-to-receiver ranges inferred from the trajectory of fragment $F_1$. The peak RCS is 61.5 dBsm,
with the noise equivalent RCS around 15 dBsm. An exponential decay fit applied over the interval
03:46:05--03:46:07 UTC gives an e-folding time of $0.69 \pm 0.13$ s (2$\sigma$), indicating that the
strongest scattering structure decays on a sub-second timescale. There are three local maxima in the
RCS that are likely to coincide with plasma irregularities from three different nearby fragments passing through the specular point
in the radar geometry.

\subsection{Radar detections relative to the optical fragments}

Panel c) of \autoref{fig:optical_radar_2x2} shows the radar detections together with the optical
fragment trajectories. The radar echoes are observed at altitudes between 55 km and 70 km and
primarily overlap with fragment $F_1$ and its child fragments. However, the Juliusruh-Bornim link
detects echoes that align better with fragment $F_2$. The clustering of detections is consistent
with the enhancement of RCS near the specular aspect angle, where the fragment velocity vector is
perpendicular to the Bragg wave vector of the transmit-receive path. Different radar links satisfy
this specular condition at different positions along the re-entry path. Scatter with non-specular
aspect angles is also detected with all links, but this scatter always has a lower signal-to-noise
ratio than the specular aspect echoes. For noisier links, the non-specular echoes are less
discernible but visible on all links, with the exception of Bornholm.

\begin{figure}[h!]
    \centering
    \includegraphics[width=\linewidth]{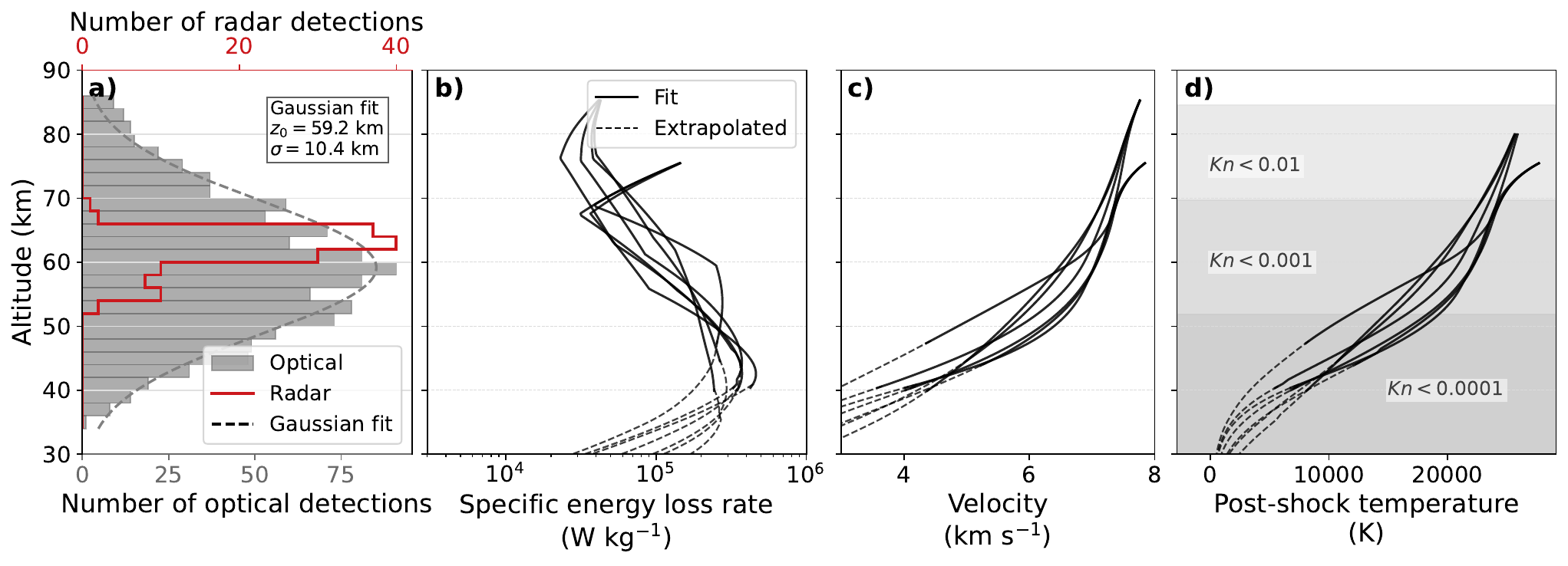}
    \caption{Summary of the altitude distribution and fitted dynamical evolution of the re-entry
             fragments. Panel a) shows the altitude histogram of all optical fragment detections
             from the triangulations together with the altitude histogram of radar detection
             heights from SIMONe. In panel a), the dashed Gaussian fit to the optical distribution
             indicates a broad maximum centered at 60\,km with a standard deviation of 10\,km,
             showing that most optical detections cluster in the same altitude region where the radar
             detections are concentrated. Panel b) shows the fitted specific energy loss rate as a function
             of altitude for the ballistic fits, with dashed segments indicating extrapolated
             trajectories below the observed range. Panel c) shows the corresponding fitted fragment
             speeds. Panel d) shows the corresponding estimated post-shock temperatures; the gray
             altitude bands mark the continuum-flow regions for 1\,m objects, using Knudsen number thresholds of 0.01, 0.001, and 0.0001.}
    \label{fig:hgt_hist}
\end{figure}

\autoref{fig:hgt_hist} summarizes the altitude distribution of fragment detections in the
optical data together with the altitude distribution of radar detections. Panel a) shows a histogram of the altitudes of all fragments detected in
cameras. The distribution indicates that fragment creation was most active between about 40 and 65
km altitude. The histogram also shows the distribution of altitudes of radar detections localized
with radar range and angle of departure. Radar detections are found between 50 and 70 km altitudes.
The dashed Gaussian fit to the optical detection heights has a maximum at 60\,km with a standard
deviation of 10\,km, quantifying the broad altitude band over which the optical fragmentation is concentrated.
Panels b)--d) show derived quantities obtained from the ballistic fits: specific kinetic-energy
loss rate per unit mass, velocity, and estimated post-shock temperature. The gray bands in panel d)
mark the Knudsen number thresholds characterizing the continuum-flow regime for a
1\,m object, assuming the trajectory of fragment $F_1$. Thresholds of $\mathrm{Kn}<0.01$, $\mathrm{Kn}<0.001$, and $\mathrm{Kn}<0.0001$ are shown. The dashed lines indicate the portions of the fits that
extrapolate outside the range of measurement points. The values of the derived parameters are
expected to be of lower quality at the edges of the measurement interval and within the
extrapolation portion of the fit.

\section{Discussion}
\label{sec:discussion}

The triangulated optical positions shown in \autoref{fig:optical_radar_2x2}a show that there are two
fragment families. Child fragments of fragment $F_1$ ($F_3$, $F_5$, $F_9$, $F_n$, $F_o$,
$F_p$, $F_r$, $F_s$, $F_w$, $F_v$, $F_x$, and $F_z$), and 
$F_2$ ($F_4$, $F_7$, $F_8$, $F_a$, $F_c$, $F_h$, $F_g$, $F_i$, $F_m$, $F_j$, $F_{\ell}$, $F_k$,
$F_t$, $F_u$, $F_e$, and $F_d$). It is not possible to determine from the measurements a point where
$F_1$ and $F_2$ are separated, as this has occurred above 85 km altitude, where the first
observation is made.

The red dashed vertical lines in panels b) and d) of \autoref{fig:optical_radar_2x2} show the
longitudes of the ground-recovered fragments in Poland reported by \cite{kruzynski2025february}.
Based on the fitted trajectories, the lower fragment $F_2$ and its child fragments are the most
likely candidates for the recovered debris, because many fragments associated with the upper
fragment $F_1$ continue further down range than the recovery locations. According to
\citep{kruzynski2025february}, the recovered debris consists of carbon fiber propellant tank parts
and thin metallic sheets. These components have relatively high area-to-mass ratios, which is
consistent with the fragments of the $F_2$ family decaying faster than fragments of the $F_1$
family, as seen in \autoref{fig:hgt_hist}c. None of the recovered fragments were associated with the
Merlin Vacuum (MVac) engine, which is expected to be more heat-tolerant and to have a lower
area-to-mass ratio than the second-stage tank structure. This supports the interpretation that
fragment $F_2$ is associated with the second-stage tank structure and fragment $F_1$
is the MVac engine.

The diameter estimates listed in \autoref{tab:ballistic_fit_summary} should be interpreted with
substantial caution. They were derived by converting the fitted ballistic coefficients into
spherical-equivalent diameters using $B=C_DA/m$, which corresponds to assuming $\rho_a'=1$ in the
dynamic model, together with an assumed drag coefficient of 2, an initial bulk density of
30~kg~m$^{-3}$ motivated by published Falcon 9 upper-stage dimensions and mass \citep{spacex_falcon_users_guide_2025}, and a terminal bulk
density of 1000~kg~m$^{-3}$ intended to represent denser compact fragments after breakup. In
reality, the fragments are not spherical, their drag coefficients are unknown and likely vary with
attitude and flow regime, the bulk density can change by orders of magnitude as cavernous
structures fragment into smaller pieces, and $\rho_a'$  can vary 10-40\% due to various atmopsheric waves \citep{fritts2003gravity,egusphere-2025-2323}. In addition,
the fitted time variable ballistic coefficient is itself only an effective parameterization of a complex
fragmentation process. The resulting diameter estimates are therefore at best order-of-magnitude
indicators rather than precise size determinations. 

The characteristics of the range-migrating radar echoes shown in \autoref{fig:rcs} differ from those
of meteor head echoes. Although the radar detections approximately track the optical fragments in
range and position, the measured Doppler shifts in \autoref{fig:rcs}b) are much smaller than
expected from the fragment motion shown in \autoref{fig:rcs}c) (tens of Hz instead of up to
$\pm$1000 Hz). This implies that the scattering plasma is not moving at the same velocity as the
object producing it, as would be expected for a meteor head echo that originates from the plasma
surrounding the ablating object. In this observation, the Doppler shift indicates motion close to
the atmospheric velocity ($<$100 m/s). A small fraction of the re-entry object's motion may still be
retained, as a slight motion along the trajectory is detectable in the Doppler shift. 

The direct optical range overlay in panel h) of \autoref{fig:snr_all_links} shows that the
triangulated $F_1$ and $F_2$ fragment families coincide with the ranges of the non-specular radar
echoes after shifting the optical fragment times later by 1.5\,s. This shift is consistent with the
observed radar echo trailing the optical fragment trajectory by about 1--2 seconds.
This indicates that the radar echo originates from turbulent plasma in the wake of the re-entry
debris. Such turbulence would take some time to develop \citep{oppenheim_ref} before significant
scattering at non-specular aspect angles can appear. The fact that the non-specular echo appears to
track the motion of the fragments without significant time or range extent suggests that this wake
plasma also decays rapidly. This decay time is approximately one second, based on the e-folding time
fit to the specular portion of the radar echo shown in \autoref{fig:rcs_peak_gate}. The turbulent
wake plasma interpretation is broadly consistent with earlier suggestions from the Mercury capsule
and Ariane 5 EPC studies \cite{lin1962radio, rubin2005airborne}. Lin also found that the brightest
radar return lagged the capsule by at least 0.6\,s, corresponding to about 4.2\,km, which is
consistent with our observation.

The re-entry vehicle produces a plasma sheath, which  surrounds the re-entry fragments.
This is evident from the loss of radio communications for re-entering spacecraft
\citep{kim2008analysis}. However, we do not observe a meteor head echo-like echo from the plasma surrounding the ablating spacecraft, despite the use of a range-Doppler matched filter that permitted large Doppler shifts. The likely reason is the width of the
plasma sheath surrounding the spacecraft. It is expected to be not much larger than the re-entry
fragment, as the mean free path at 60 km is only around 1 mm. Thus, the RCS of the head echo is not
expected to be much larger than the size of the parent fragment that produces the plasma sheath
surrounding the spacecraft. The Falcon 9 second stage is approximately 15 m long and 3.66 m in
diameter, excluding the payload adapter and fairing \citep{spacex_falcon_users_guide_2025}. Even
taking the circular cross section of a 3.66 m diameter stage as a crude geometric upper bound gives
only about 10 m$^2$, which is still well below the 100 m$^2$ detection limit of the SIMONe radar
system for the location where the radar echoes were detected. However, once the plasma turbulence within the wake develops sufficiently, the RCS is increased above the detection limit. Had the re-entry debris been tracked with a more powerful radar, such as an MST radar, it is likely that the objects could have been tracked using direct scatter from the fragments themselves, without a significant increase in RCS due to the plasma sheath.

While the optical data does not allow absolute intensity of optical emissions to be estimated,
visually the main fragment $F_1$ is significantly brighter than fragment $F_2$, as shown in
\autoref{fig:grid_2x3}. This indicates higher temperature and hence higher plasma production. This
is consistent with radar signatures primarily being associated with fragment $F_1$, as shown in
\autoref{fig:optical_radar_2x2}. As discussed earlier, this fragment is suspected to be associated
with the MVac engine, which implies that the composition and the ballistic coefficient of the
re-entry object could play an important role in the radar detectability of re-entry debris with
radar.


The enhancement of RCS at specular geometry, shown in \autoref{fig:rcs}a) together with the small
observed Doppler shifts in \autoref{fig:rcs}b), has a resemblance with overdense specular meteor
trail echoes \citep{mckinley1961meteor,poulter1977radiowave}. Such an enhancement was also reported
by \cite{lin1962radio} and \cite{rubin2005airborne}. This enhancement at specular aspect angle
suggests that there is a narrow, elongated plasma structure within the 1-2 km long Fresnel zone
along the path of the re-entry debris, which contributes coherently to the formation of a radar
echo, with a cross-track extent less than the radar wavelength of 9.2 meters. The echo power decay
time constant of $0.7 \pm 0.1$ s is significantly shorter than typical e-folding timescales of
overdense specular meteor trails. One explanation is that D-region recombination ion chemistry is
the main controlling factor for trail decay in the 55-65 km altitude range, instead of the much
slower ambipolar diffusion, which becomes more effective than ion chemistry at higher altitudes
\citep{turunen1996incoherent,younger2014effects}. It is expected that the chemical composition of
the ablated material will also influence the ion chemistry.

The production of plasma around re-entering orbital objects differs from that of meteors. The
re-entry debris observed in this study has velocities of 6--7 km\,s$^{-1}$ at the time when radar
echoes are observed, as shown in \autoref{fig:hgt_hist}. Meteor head echoes are rarely detected at
such low velocities because they fall below the impact-ionization threshold that governs plasma
formation for faster meteoroids \citep{vondrak2008chemical}. This implies that atoms ablated from
the heated spacecraft would not ionize efficiently through collisions with atmospheric molecules
alone. However, strong plasma echoes were still detected with very large RCS values. The ionization
mechanism for the observed plasma is the high temperature shock front. At around 60 km altitude, the
Knudsen number, which is the ratio of the mean free path to the object scale size $\mathrm{Kn} =
\lambda_{\mathrm{MFP}}/L$ is rather low. Based on Table \ref{tab:ballistic_fit_summary}, the fragments detected optically with the cameras are estimated to have diameters in the 0.5-5 meter range with Knudsen numbers $0.0002<\mathrm{Kn}<0.002$ in the continuum flow regime, which is a necessary condition for a ionizing high temperature shock front to
form \citep{grossir2018high,esposito2024non}. \autoref{fig:hgt_hist}d illustrates the estimated post-shock gas temperature using \autoref{eq:shock_temp}. Limiting estimations to $\mathrm{Kn}<0.005$, the temperatures indicate that significant plasma production can occur due to the shock above 40 km. Even though it is outside the scope of this study to simulate the shock plasma production precisely, typical relations such as the Saha ionization equation, show that significant thermal ionization begins at 10-20 kK \citep{heSimplifiedMethodCalculating2021}.

The mean free path
at 60 km is around 1 mm, so objects larger than 1 cm could potentially produce an ionizing
shock front at this altitude \citep{grossir2018high}. Applying equation \ref{eq:dynamic_model} to modeling of orbital circular orbit re-entry indicates
that objects smaller than 1 cm would decelerate significantly due to atmospheric drag before
reaching an altitude where the slip-flow or continuum-flow regime necessary for plasma formation
exists. Thus, objects smaller than this are unlikely to create sufficient plasma density to allow
detection by specular meteor radars. The fact that we only see signatures of four fragments in the
specular and non-specular radar echoes, indicates that there is a lower size threshold for sufficient
plasma production to be detectable by specular meteor radar systems, and this threshold for the SIMONe radar system is around 1 meter based on the number of fragments detected with the radar, the diameters estimated based on the ballistic fits, and the size of the Falcon second stage. More detailed modeling of the plasma production and radar scattering would be needed to
determine the expected radar signatures for different fragment sizes and to constrain the size
threshold for radar detectability with different radar systems.


Understanding the differences between the re-entry behaviour of meteors and the slower, uncontrolled
re-entry of the Falcon 9 is essential for quantifying the impact of space debris re-entry on the
middle atmosphere. As noted in the introduction, \citet{wing2025measurement} observed a lithium
plume which formed as the Falcon 9 hull heated above 933 K upon re-entry at an altitude near 100 km.
In that study, the authors used a simplified model for the ablation of aluminium rockets, which
neglected rocket body fragmentation, shock-front formation and ionisation. The authors of the study
also acknowledged in their discussion that the re-entry trajectory of the Falcon 9, provided by ESA,
was the largest source of uncertainty in linking the detection of a mesospheric lithium plume to the
Falcon 9. The use of camera and radar networks to accurately characterise the re-entry trajectory,
including plasma generation of debris fragments, would address the weakest aspect of that study and
could provide a strong contribution for future studies of its kind and potential environmental
monitoring activities.
The Gaussian fit to the optical detection-height distribution in \autoref{fig:hgt_hist}a, centered
at 60\,km with $\sigma=10$\,km, provides an observational summary of where fragmentation was
most visible. While this is not a direct measurement of mass deposition, the distribution of optical
signatures is expected to be correlated with it and therefore offers a useful empirical constraint
for comparisons with the altitude range of material deposition discussed by \citet{murphy2023metals}.

\conclusions 
\label{sec:conclusion}

This study provides a dataset of 30 fragment trajectories derived from stereoscopic triangulation of
the February 19th 2025 Falcon 9 upper-stage re-entry. During the most intense phase of the breakup,
between 55 km and 70 km altitude, not every visible fragment could be identified because of the
large number of simultaneously visible objects. Nevertheless, the reconstructed trajectories show
that the lower fragment family $F_2$ is likely the propellant tank assembly and the source of the recovered ground debris,
whereas the upper fragment family $F_1$ is consistent with a denser MVac engine. Ground recovered fragments were associated with fragments that broke off of $F_2$, whereas radar signatures were linked primarily to the optically brighter
fragment $F_1$.

The dynamic fit of the trajectories indicates that the largest kinetic-energy loss occurs
at 30--70 km altitude, which is also where the largest number of optical fragments and the radar
echoes were observed. Modeling and ablation chemistry studies suggest that re-entering space
materials deposit material into this altitude range \citep{murphy2023metals},
which agrees with the observations. Without additional re-entry modeling that uses a more detailed
engineering model, such as SCARAB \citep{lips2004spacecraft}, it is not possible to determine
precisely how the mass deposition is distributed with altitude. Comparing the observations in this
study with such a model would therefore be an valuable next step. It is possible that the radar
echoes can help constrain such models, because they provide independent information on the altitude
range where plasma production and mass loss occur. The optical observations, even though not
absolutely calibrated, could also yield useful constraints on radiation losses during re-entry,
allowing better constraints on heating rates of the spacecraft. Further, co-location with lidars
would allow for additional constraints by simultaneously evaluating the re-entry trajectory,
geophysical dynamics as well as composition for both mesospheric metals produced during ablation as
well as potential stratospheric aerosols which may be produced during fragmentation.
The fitted optical altitude distribution, with a peak at 60\,km and a standard deviation of
10\,km, provides a compact observational benchmark for that comparison. Although it is not a
direct measurement of mass deposition, the optical-signature distribution is expected to be
correlated with the altitude range where deposition is strongest.

Two distinct types of radar echoes were observed: 1) radar echoes at non-specular aspect angles with
30-40 dBsm RCS, 2) echoes with enhanced RCS at specular aspect angles, with up to 60 dBsm RCS. The
suggested mechanism of ionization is a superheated plasma created within the hypervelocity shock
front. The non-specular aspect angle echoes are explained as ionized wake turbulence, and the
specular echoes are explained as a structure similar to overdense specular meteor trail echoes
\citep{mckinley1961meteor,poulter1977radiowave}. 

As shown in this study, fairly low power
all sky meteor radar systems are able to observe re-entry plasma with 
significant RCS enhancements due to plasma produced. These echoes are likely
 to be observed in the region where slip flow or continuum flow conditions are met,
  which is expected to be below 
 60 km altitude for objects larger than a few centimeters in size. There are many similar 
 sytems around the world, each 
 with a collecting area in the order of 100000 km$^2$. These systems could potentially 
 contribute in the future to monitoring
  the re-entry of space debris. It is possible that more sensitive radars could detect plasma trails of smaller re-entry objects than shown in this study, which are expected to produce ionization when they are in the continuum flow regime. 
  
  In this study, the fragments that produced radar echoes are estimated to be in the 0.5--5 m size
  range. Objects of this size are already tracked by the U.S. Space Surveillance Network, so
  radar observations are not expected to provide substantial additional information about the number
  of re-entering objects. However, meteor radar and optical observations can provide useful information
  about plasma production and mass deposition during re-entry, which is important both for
  understanding the impact of space-debris re-entry on the atmosphere and for improving
  ground-impact hazard assessment.

\dataavailability{The optical and radar dataset associated with this study is publicly available \citep{vierinen_2026_20070800}. ERA5 pressure-level reanalysis data used for the wind correction were obtained from the Copernicus Climate Change Service Climate Data Store \citep{hersbach2023era5_pressure_levels}.}

\authorcontribution{JV led the analysis and manuscript preparation. DKn contributed optical data analysis. JLC, GB, DH, MC, NP, TR, RW, and KSO contributed radar observations, analysis, and interpretation. BG contributed optical calibration and interpretation. DKa contributed interpretations of the re-entry. All authors contributed to discussion of the results and manuscript revision.}

\competinginterests{The authors declare that they have no conflict of interest.}

\begin{acknowledgements} Part of the work presented here is based on data from the AllSky7 camera
network (\href{https://allsky7.net}{allsky7.net}). We thank the camera operators of the AllSky7 network for providing optical observations used in this study: John Downing (AMS101), Jörg Strunk (AMS21), Sebastian Schmidt and Johannes Schultz (AMS216), Jochen Schönfelder (AMS233), Sirko Molau (AMS35), Gerhard Lehmann (AMS55), Martin Fiedler (AMS65), Bernd Klemt (AMS76), Björn Poppe and Theresa Ott (AMS95), Sirko Molau (AMS16), Julia Lanz-Kröchert (AMS213), André Knöfel (AMS22), Stiftung Planetarium Berlin (AMS238), Peter Kroll, Mario Ennes, and Thomas Müller (AMS52), Maciej Libert and Jan-Gerd Mess (AMS62), Luc Bastiaens (AMS67), and Peter Lindner (AMS88). This work contains modified Copernicus Climate Change Service
information (2025). Neither the European Commission nor ECMWF is responsible for any use that may be
made of the Copernicus information or data it contains. Large language models were used for grammar
and programming assistance only. JLC and MC acknowledge support from the European Office of AFOSR 717
(EOARD) grant FA8655-23-1-7017. \end{acknowledgements}

\bibliographystyle{copernicus}
\bibliography{refrences.bib}

\end{document}